\documentclass[conference]{IEEEtran}
\IEEEoverridecommandlockouts
\usepackage{cite}
\usepackage{amsmath,amssymb,amsfonts}
\usepackage{algorithmic}
\usepackage{graphicx}
\usepackage{textcomp}
\usepackage{xcolor}
\usepackage{listings} 
\usepackage{cleveref}
\def\BibTeX{{\rm B\kern-.05em{\sc i\kern-.025em b}\kern-.08em
    T\kern-.1667em\lower.7ex\hbox{E}\kern-.125emX}}
\begin{document}

\title{Enabling Cross-Language Data Integration and Scalable Analytics in Decentralized Finance}

\author{\IEEEauthorblockN{Conor Flynn}
\IEEEauthorblockA{\textit{Electrical and Computer Systems Engineering} \\
\textit{Rensselaer Polytechnic Institute}\\
Troy, United States of America\\
flynnc3@rpi.edu}
\and
\IEEEauthorblockN{Kristin P. Bennett}
\IEEEauthorblockA{\textit{Mathematical and Computer Sciences} \\
\textit{Rensselaer Polytechnic Institute}\\
Troy, United States of America\\
bennek@rpi.edu}
\and
\IEEEauthorblockN{John S. Erickson}
\IEEEauthorblockA{\textit{Future of Computing Institute} \\
\textit{Rensselaer Polytechnic Institute}\\
Troy, United States of America\\
erickj4@rpi.edu}
\and
\IEEEauthorblockN{Aaron Green}
\IEEEauthorblockA{\textit{Mathematical Sciences} \\
\textit{Rensselaer Polytechnic Institute}\\
Troy, United States of America\\
greena12@rpi.edu}
\and
\IEEEauthorblockN{Oshani Seneviratne}
\IEEEauthorblockA{\textit{Computer Science} \\
\textit{Rensselaer Polytechnic Institute}\\
Troy, United States of America\\
senevo@rpi.edu}
}

\maketitle

\begin{abstract}
    With the agile development process of most academic and corporate entities, designing a robust computational back-end system that can support their ever-changing data needs is a constantly evolving challenge. We propose the implementation of a data and language-agnostic system design that handles different data schemes and sources while subsequently providing researchers and developers a way to connect to it that is supported by a vast majority of programming languages. To validate the efficacy of a system with this proposed architecture, we integrate various data sources throughout the decentralized finance (DeFi) space, specifically from DeFi lending protocols, retrieving tens of millions of data points to perform analytics through this system. We then access and process the retrieved data through several different programming languages (R-Lang, Python, and Java). Finally, we analyze the performance of the proposed architecture in relation to other high-performance systems and explore how this system performs under a high computational load.
\end{abstract}


\section{Introduction}

    Designing a robust computational back-end system poses a series of challenges many software engineers face daily. Languages all offer a variety of functionality, each with its benefits and drawbacks, causing developers to select a primary language based on the specifications of the project \cite{goosen2007choosing}. However, the selection of this language is often pivotal, as it determines not only the set of tools at a developer's disposal but often the ability to scale a system outside of its original confines.

    Applications revolving around the transferring, storing, and processing of big data have been evermore prevalent as well. Although the retrieval of the data may be standardized to a certain degree through means such as OpenAPI standardization and Javascript Object Notation (JSON) formatting, the data being transmitted is not \cite{openapi} \cite{bray2014javascript}. Often this passed data has nested values (in the case of JSON, such as having JSON objects and JSON arrays), different primitive object types, and varying subsequent request mannerisms, all of which account for increased system complexity and the potential for over-engineering solutions \cite{oracle2023jsonobject} \cite{oracle2023jsonarray} \cite{oracle2023datatypes} \cite{alexander2008overengineering}.

    From this, we define a problem that exists in both educational and business domains, being the lack of heterogeneity in data sets as well as the inability to expand upon a system in a practical manner. In this paper, we discuss the implementation and deployment of a system that accounts for these discrepancies between data sources and languages, resulting in a high-performance back-end solution for standardizing data transmission and storage between multiple data sources and user applications. We also analyze the underlying architecture of the system and how it can be independently utilized to create scalable solutions, how it is used in our application, and what it means for the system's architecture moving forward.

    Based on these claims, we propose a solution called the "Decentralized Finance Data Engine," (DFDE) which utilizes the proposed system architecture. Decentralized Finance (DeFi) is a "new breed of consumer-facing financial applications composed as smart contracts, deployed on permissionless blockchain technologies" \cite{jensen2021defi}. The architecture includes features such as multi-tenant architecture with asynchronous actions, programming language agnostic connections, and support for multiple data sources within the DeFi space focusing on different lending protocols such as the AAVE protocol \cite{chong2006multi} \cite{aavewebsite}. This implementation is then used in an educational lab setting, where researchers can easily pull and manipulate this data to expedite their work.

    To introduce this work, we first cover related works with similar purposes followed by the relevancy of such a system and why it is beneficial to these different entities. Following this, we introduce the underlying architecture the DFDE is based on and analyze the performance of the design. Next we review proposed solutions for the internal and external connections such an implementation requires. Finally, we review the DFDE's architecture as a whole, how the different processes within it interact, and the applications it is used for.

\section{Related Work}

    Although this system proposes several improvements to the interoperability of systems and their transfer of data both internally and externally, one should recognize the existing developments in this area; primarily concerning the Application Programming Interface (API) specifications/protocols \cite{altex2022api}.
        
    When regarding APIs and the manners to transfer data, almost all designs will fall under a specific API specification using either Remote Procedure Call (RPC), Service Object Access Protocol (SOAP), Representational State Transfer (REST), gRPC, or GraphQL. In this paper, we explore implementations using REST and GraphQL protocols.
    
    A REST API refers to an API that follows six architectural constraints: uniform interface, stateless, cacheable, client-server, layered system, and code on demand \cite{fielding2000architectural}. REST APIs are commonly used for create, read, update, and delete (CRUD) operations on remote systems, which is why they are primarily used for exposing server data to external clients \cite{aws2023rpc}.

    GraphQL acts as a query language for APIs \cite{altex2022api}. It allows users to specify the exact data needed from the source and then retrieve it in one call to the API. This approach aims to save both time and computational resources for the client and server and is adopted in scenarios requiring large amounts of data to be transferred.

    % One commercialized related project is called RapidAPI \cite{rapidql2023docs}. 
    RapidAPI \cite{rapidql2023docs} aims to generalize API calls through the use of a query language, similar to that of SQL and GraphQL, for the easy retrieval of data from an external source \cite{stonebraker2010sql}. Similar to the proposed implementation, RapidAPI allows for MongoDB integration and connection as well as the caching of previously called data. Although very similar in usage to the proposed sample application, referenced in \cref{sec:dfde}, there are some key differences. 
    
    The primary difference is how the system communicates with end users. RapidAPI features a query language for users to specify explicit instructions which their system interprets \cite{rapidql2023connection}. Although great for the necessary applications, developers cannot easily expand the system or use it for other capabilities, which are extrapolated upon in the Routing Architecture section. Its usage also revolves around a query language, which may not be language agnostic as using something low-level such as Sockets \cite{kalita2014socket}. Finally, their requesting system is much more specific than the proposed implementation. This has benefits, such as being able to request data in a more specific manner, while also having its limitations, such as more specifications being required when executing a call. It cannot also perform recursive API calls automatically, requiring that handling be done on the user's end.

\section{Relevancy}

    Before analyzing the underlying architecture of the DFDE and its processes, one must understand the relevancy of such a system and the problems to which it can be applied. Primarily, the DFDE aims to resolve problems in relation to API design; however, these questions can be expanded upon to any form of data transfer. As Lindman states regarding business, architecture, process, and organization (BAPO) perspectives:
    
    \begin{quote}
        The main concern of this perspective is to investigate the technical issues associated with API design and development, similar to the API layer presented but from a broader perspective (architecture of the interacting systems as opposed to API design). Questions to ask include: How does one manage API versioning (e.g., side-by-side deployment of different versions)? How does one design APIs for an extension? How does one check the backward compatibility of APIs between different versions? \cite{lindman2020APIs}
    \end{quote}
    
    This gives insight into the troubles faced by researchers regarding designing APIs (or any outward-facing programs). However, the DFDE offers a solution to many of these posed questions.
    
    Firstly regarding versioning and backward compatibility, the DFDE does inherently hold backward compatibility functionalities so long as the underlying Router structure and connection mechanisms are not modified. Processes may change. However, unlike APIs, where the structuring of data, endpoint connection, and parameters may change, the DFDE's internal calling to external APIs will always remain consistent. The only inconsistency that may be present is an external data provider's API connection may change, causing the DFDE's internal storage of data and the outward-facing connection to be incorrect, requiring manual maintenance by the user of the DFDE to fix the connection. This makes it so researchers have to make minimal changes to their existing applications using the DFDE, allowing for minimal obstruction to the lab's workflow with any sudden pivots in the focal research topic.
    
    Similarly, extensions to the DFDE can be made easy thanks to the Router design. With the only internal modification requirement being connecting the new Router(s) to the rest of the system, there is not much else a programmer has to do to expand upon the DFDE. To access the new Router(s) and sub-tag(s), users simply have to specify their location when submitting requests. All internal processes are modular, making future updates easy. With this, any students working on a project irrespective of being an original developer can easily add their own components to the DFDE.
    
    Finally, due to the data-agnostic designs of the DFDE, new data sets can be easily parsed, stored, and analyzed by researchers. The language-agnostic design also allows students to work in their preferred languages, increasing their productivity and the efficacy of their work while helping to reduce the friction in learning a new language \cite{shrestha2020here}. This allows for the usage of such a system in a classroom setting as is currently being done with the proposed application of the DFDE, allowing students to explore material outside of the confines of the course.
    
\section{Routing Architecture}

    This section describes the internal data routing architecture of the DFDE, on which all processes are inherited.
    
    \subsection{Overview}
    
\begin{figure}[ht]
        \centering
        \includegraphics[scale=0.5]{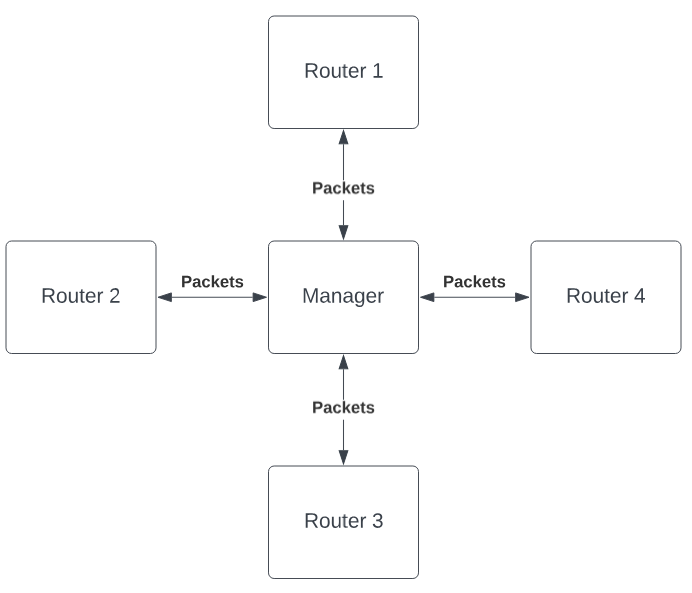}
        \caption{Routing Architecture}
        \label{router-overview}
    \end{figure}

    To design a language and data-agnostic system, a highly versatile and robust architecture was needed. Therefore the proposed architecture is modeled after a network consisting of Routers and Packets, with different Packets of data being able to be transmitted freely between the connected Routers \cite{oracle2014}. We use \cref{router-overview} to give a visual of the routing architecture.
    
    \subsection{Managers}
    
        Managers are the primary link between all Routers, with each Router sharing the same Manager. All Routers within a manager are stored using a HashMap, with the key being the unique "Tag" of a Router and the value being the Router object itself \cite{oracle2023hashmap}. This way, there is a fast connection when two Routers are trying to communicate, as a HashMap uses constant time when accessing information \cite{bajracharyareview}. 
    
    Managers also are "intelligent" objects recognizing when two groups of Routers, each with a different Manager, connect. In such cases, they will override all of the Routers to have the same Manager so that constant run-time is retained. Managers also handle all edge case connections, such as duplicate connections, disconnections, and threaded processes \cite{holub2000taming}.
    
    \subsection{Routers}
    
        Routers are the interactive endpoints for all processes. They contain standardized functions that allow for easy connection and communication with other Routers. Each Router has a "Tag," which is a unique String identifier so a Manager can recognize it and properly route Packets to it \cite{oracle2023strings}. Routers also consist of processes, all denoted by a unique String called a "Sub-Tag" for proper Packet transit within the Router. Each process is a defined method that returns a Response containing specific information regarding the call \cite{oracle2023methods}.
    
        Similar to Managers, Routers also contain "intelligent" capabilities such as the ability to: recognize specifically formatted methods as process functions (with unique identifiers), send Packets to these requested functions, and recognize any formatting errors when defining these functions during the initial run-time compilation. To accomplish some of these functionalities, the architecture uses a feature of the Java language called Reflection, which allows for the Java program to analyze the code on which it is running and perform actions accordingly \cite{forman2005java,ronmamo2021reflections}.
    
    \subsection{Packets}
    
        Packets are lightweight objects comprised of four main pieces of information: the sender, the receiver (referred to as the "Tag"), the process (referred to as the "Sub-Tag"), and all internal data \cite{flynn2023packets}. The sender, Tag, and Sub-Tag are all listed as Strings, whereas the data is stored as a HashMap (with both the key and values being Strings) \cite{oracle2023strings} \cite{oracle2023hashmap}.
    
        Each call to a Router's process then returns a Response. Responses contain information regarding the call's outcome and are comprised of a response code, a message, and optional return data. The response code is a predefined integer representing a specific outcome (commonly following the OpenAPI HTTP Status Code standard)\cite{oracle2023datatypes, openAPI2023httpstatuscodes}. The message is also predefined and is a String corresponding to the returned response code, giving more insight into the response of the code. Finally, the data refers to any optional data requested by the Packet that may need to be returned. Although the data is returned as a String, it can be cast into any needed data type by the requesting process \cite{Baeldung_2021}.
    
    \subsection{Review}
    
        When using this architecture in the implementation and design of a large system, several benefits and drawbacks can be found, with their severity on the implementation varying based on the requirements.
    
        Benefits include the easy scalability of the system, the generalized data flow, and the performance, later discussed in the \cref{sec:performanceAnalysis} section. When designing new systems, developers may often specify the system to the current needs, not foreseeing future updates and improvements. This often creates a very rigid programming schema, requiring developers to spend more time when making these updates. The architecture proposed in this paper solves this problem, allowing for the easy implementation and addition of new components without requiring explicit modification to existing regions. Similarly, this generalized data flow allows for easy communication between different sections of the system, making any changes easy to implement. Added latency is also very minimal, making it a very strong contender on time-sensitive programs.
    
        The primary drawback of this architecture is the generalized data flow, specifically the casting required to transmit information. Since all data is passed as a string, it may often be challenging to communicate complex objects between portions of the system. A solution to this could be to create the Packets using an \texttt{Object} value rather than a \texttt{String}. However, further checks would need to be done to ensure the safety of casting the object \cite{fruit1993object}.

\section{Performance Analysis}\label{sec:performanceAnalysis}

    This section discusses the performance of the proposed routing architecture and how one should interpret the results.
    
    \subsection{Experimental Setup}
    
        \textbf{Hardware:} The environment used contains a single Intel(R) i9-12900KF processor with 16 cores, in which each core runs at 3.187 GHz and 64GB of memory.
    
        \textbf{Software:} The environment is using Microsoft Windows 10 Pro version 10.0.19045 Build 19045 as an operating system. The programming language Java is used to run the tests, using Java 20.0.1, build 20.0.1+9-29.
    
    \subsection{Overview}
    
        To analyze the performance of the proposed architecture, we test the rate at which two Routers can submit packets of data to one another. To accomplish this, we send a packet containing some information to a sample protocol on a connected Router. We then analyze the time it takes for that packet to reach the Router and for a response to be received. Once received, we send another packet and repeat the cycle for 10 seconds until the run is up. We then average the packets sent per second over these 10 seconds to determine the performance.
    
        We also analyze the maximum capacity of the architecture through threading, calculating how many packets can be sent asynchronously. To understand how threading affects performance, we begin with a singularly threaded test and increase the number of threads by a factor of 10 until we reach 90 concurrent threads.
    
    \subsection{Analysis}
    
\begin{figure}[ht]
            \centering
            \includegraphics[scale=0.45]{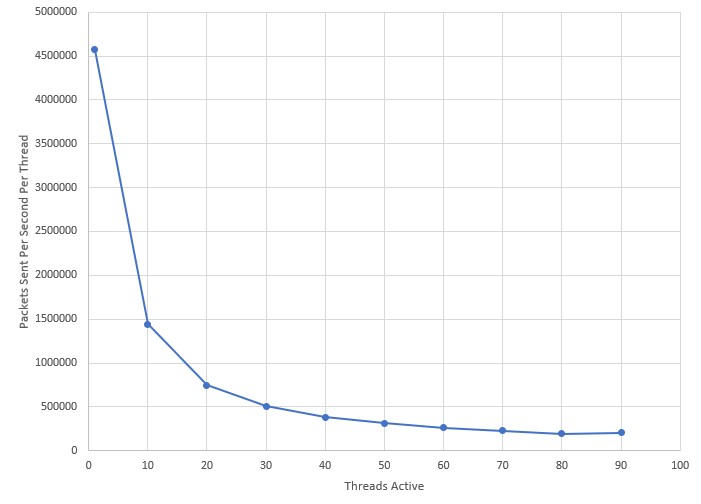}
            \caption{Packets Sent Per Second Per Thread}
            \label{routing-threads1}
        \end{figure}

        We first review the performance of each thread regarding the number of packets being sent. As seen by \cref{routing-threads1}. There is a significant drop in performance between 1 and 10 threads, which is to be expected. However, this degradation in performance becomes more minimal as more threads are used. From this, we can determine the time it takes for a packet to be sent and received by dividing the number of nanoseconds per second (1 billion) by the number of packets sent per second. This tells us that it takes 218.78ns to send a packet within a single thread versus 4835.81ns to send a packet when 90 threads are running concurrently.
    
\begin{figure}[ht]
            \centering
            \includegraphics[scale=0.42]{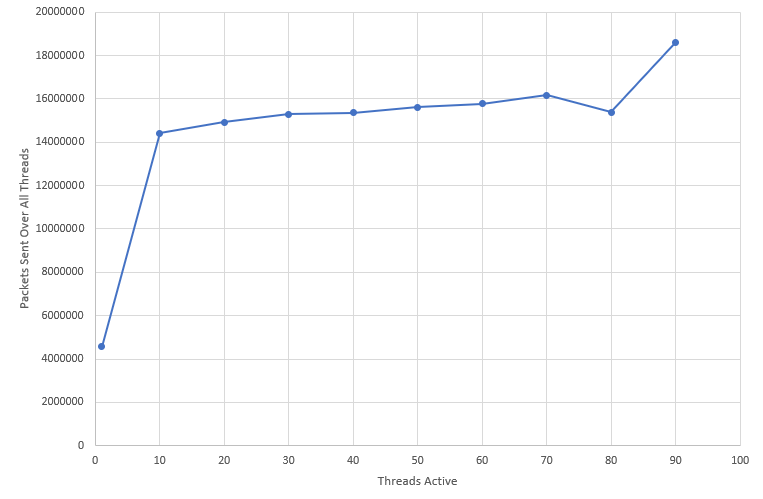}
            \caption{Packets Sent Per Second Across All Active Threads}
            \label{routing-threads2}
        \end{figure}

        Although this drop in performance is very large, we can also extrapolate the total number of packets being sent per second. To do this, we take the current number of packets sent per second and multiply it by the number of threads actively running. As seen in \cref{routing-threads2}, although the speed at which packets are sent is greatly reduced, the number of overall packets being sent is greatly increased, with 90 threads handling over 4 times the number of packets as a single thread.
    
        From this, we can see the trade-off of performance versus capacity. Should a system need fast interactions between processes, it may opt to limit the number of threads so that performance is retained. Likewise, if an application relies on transmitting large quantities of information with minimal emphasis on time, it may choose to operate with a large number of threads.

\section{Internal Storage}

    This section covers all internal architecture relating to the processing, storage, and retrieval of all agnostic data handled by the DFDE.
    
    \subsection{Data Schema}
    
        Standardizing a database with no context of the incoming data is certainly challenging. To solve this problem, we rely on the preliminary condition of the data being structured properly before the DFDE interprets it. The required format is JSON, a lightweight, text-based, language-independent data interchange format \cite{bray2014javascript}. From this, it can be easily converted into Binary JSON (BSON) so the MongoDB can properly process the data points \cite{mongo2023JSON}. Through the use of recursion, we can easily extract information embedded within the calls and use it if necessary \cite{dijkstra2022recursive}. Use cases include recursive calls to endpoints, extracting storage information, and fast data filtering.
    
    \subsection{Collection Management}
    
        To house the formatted data points, MongoDB relies on Collections \cite{mongo2023Collections}. Collections are groups of data points stored under a common name, which a database can easily access. This includes two primary components: the request name and date (should the data be dated). Request names are required to be unique on initialization, with the DFDE checking each request for uniqueness. This unique name allows for consistency within the DFDE, allowing for calls to be validated if the data has already been received or not. For dated data, the naming schema changes slightly. Rather than just being the unique name passed, it will also include the date to be stored in the format: \texttt{name-yyyy-MM-dd} with the year, month, and day following the standardized "Formatting and parsing date-time patterns" \cite{ibm2023datetime}. Although this schema is trivial, it allows for a variety of functionality which will be further expanded upon in later sections.
    
    \subsection{Review}
    
        There are many benefits to using a data-agnostic system such as the DFDE, as described above, which include easy accessibility, reduction of latency, and credential obfuscation. Easy accessibility refers to the availability of the data to both users operating with the external protocols mentioned later as well as to users directly interacting with the MongoDB \cite{mongodb2023docs}. Reduced latency refers to the reduction of calls to external data sources since called data is cached and stored for subsequent calls. Finally, credential obfuscation means that one user can request data using an API key and subsequent calls to retrieve the data without requiring the key, as the data is already cached and stored properly.

\section{External Data Connections}

    This section covers all external connections to data sources outside the scope of the DFDE. This includes but is not limited to, external REST APIs, WebSockets, and Web3 connections \cite{rodriguez2016rest, fette2011websocket, yang2023web3}.
    
    \subsection{Abstract Connections}
    
        Designing a system to account for every existing type of available endpoint is nearly impossible. Between different endpoint schemes used by companies, the types of connections offered (such as REST APIs and WebSockets), and the data returned by these endpoints, a system would have to be needlessly over-engineered, defeating the purpose of a high-frequency data transferring system.
    
        To counter this issue, we implemented generic and commonly used protocols while also leaving the DFDE open to abstraction for further development. This way, should the integrated solutions not be suited for a given project, developers can easily integrate their own abstract endpoints with the required specifications.
    
    \subsection{Configuration}
    
        Before designing a system to cater to these different data sources, we first need to comprise a way for the DFDE to easily interpret new endpoints. This can be achieved through the usage of properties files, which are files that contain specific key-value pairs able to be read by the DFDE. By creating a directory that houses these different configuration files, the DFDE can load them on initialization and recognize/handle incoming requests.
    
    \subsection{REST APIs}
    
        The primary generalized architecture supports external calls to REST APIs \cite{rodriguez2016rest}. REST APIs consist of several different properties, primarily: a Uniform Resource Locator (URL), properties defined after the URL, and headers to specify specific underlying properties not passed within the URL \cite{berners1994uniform}. There are also edge cases, such as URLs having a unique path differing from the one specified, which the DFDE must also account for. To handle these different cases and information required, we have variables within the properties files such as \texttt{url.base}, \texttt{url.properties}, and \texttt{url.headers} which will properly translate the passed requests into the proper REST API calls necessary to retrieve information.
    
        Other information can be necessary, such as retrieving dated data. Rather than having users make individual calls to retrieve each date, we can use a recursive function to extrapolate every date and make each call on behalf of the user. For this, we have key-value pairs that define the variables used for setting the properties controlling the dates, such as \texttt{date.start} and \texttt{date.end} as well as ones defining how the date is formatted \texttt{date.format} using the same schema as the collection management system \cite{ibm2023datetime}.
    
        Finally, the DFDE has to account for recursive calls. This is because most external systems have a limit as to how many data points a single REST API call can return. To determine if a recursive call has reached the end of the data (stored within the given dates or endpoint), we use a limiter. If the number of data points returned is below the limit defined, we deem the call finished allowing it to move to the next call. While designing the DFDE, we identified four main types of recursive calls: Single, URL, Incremental, and Static.
    
        \subsubsection{Single}
    
            A single call is the most basic type of call. It tells the DFDE that no recursion is required and there is only one request needed to retrieve all the data. This is commonly used for retrieving static data sets containing reference data.
    
        \subsubsection{URL}
    
            A URL call states that there is an iterative URL returned with the data that the user should use to make the next call. For this type of call, the DFDE just needs to know where the new URL is located in the returned data, and it will be able to substitute it to make the next call.
    
        \subsubsection{Incremental}
    
            An Incremental call tells the DFDE that there is a variable that needs to be increased to retrieve the next group of data. Commonly used for endpoints that have "Pages" of data, and the page needs to be increased by one each call to obtain the next set of data.
    
        \subsubsection{Static}
    
            A static call refers to a data point that is returned inside the passed call that needs to replace a parameter in the URL. This is typically a timestamp or index where by passing the timestamp or index, the external data source will return a certain number of data points after that value.
    
            Due to the probability of more niche cases than these, the DFDE allows for the abstract configuration of external calls, as mentioned previously.
    
    \subsection{WebSockets}
    
        WebSockets refer to a form of continuous communication and data flow, where one system acts as a client and the other as a server \cite{fette2011websocket}. This form of constant communication is seen primarily in live data feed applications. To integrate connections such as these into the DFDE, all that is necessary is an establishment of the connection and the usage of the \texttt{PUSH} call as defined by the \texttt{LocalStreamHandler}. 
    
    \subsection{Web3}
    
        Designing a Web3 connection is most likely the most specialized type of connection supported by the DFDE, as there are many ways to connect to decentralized applications. Later in this paper, we will explore connections to AAVE and The Graph, both of which are Web3 applications. However, should users want to obtain other Web3 data, it will most likely require them to develop a new connection utilizing the abstract connection types discussed previously \cite{Frangella_Herskind_2022, thegraph2023}.

\section{Language Agnostic Connections}

    This section contains information regarding the programming language features required for connection, the steps to connect, and the embedded multi-tenant architecture of the DFDE \cite{chong2006multi}.
    
    \subsection{Language Feature Requirements}
    
        Outside of common object-oriented programming functionalities, the only language requirement is the support of Socket connections. \cite{stroustrup1988object} \cite{donahoo2009tcp}. Since Sockets are considered a basic programming feature, most modern generic programming languages will support their usage \cite{kalita2014socket}.
    
        Sockets were selected as the primary method of choice for transmitting data for a few reasons. Commonly referred to as WebSockets, this form of communication "provides a full-duplex, bidirectional communication channel that operates through a single Socket over the Web" \cite{pimentel2012communicating}. This gives them several large advantages over REST API calls with functionalities such as reading and writing capabilities, non-capped data transfer (typically capped at 2MB for common server instances), and a constant line of communication \cite{apache2012tomcat}. The usage of all of these functions will be explored in later sections.
    
    \subsection{Initializing Connections}
    
        There are a few steps required to connect to the DFDE before requesting and retrieving data.
    
        \subsubsection{Destination Key Retrieval}
        
            Upon a user initiating a connection with the DFDE, the DFDE will submit a one-line response to the Socket containing a unique key. This key, referred to as the \texttt{destination-key}, is the unique identifier of the application. It tells the DFDE where to send data that is requested. When a user requests data from the DFDE, they are also required to submit their \texttt{destination-key} so that the DFDE can properly transmit their data. One benefit of this is an application can asynchronously request data to be sent to other applications by sharing each other's keys.
    
        \subsubsection{Sending a Request}
        
            Due to the way requests are processed, the DFDE has strict syntactical requirements for the formatting of requests. Since the DFDE utilizes the Router architecture previously discussed, a request can be made to any Router's tag and sub-tag. To send requests to the DFDE, we utilize the tag \texttt{SRC} (the stream registry controller) and the sub-tag \texttt{RQST} (the request protocol). 
        
            Next, we specify the required details relating to the call. To handle these different parameters, the DFDE has a specified delimiter (defaulted to \texttt{\&\&\&}) which is used to isolate these values. For calls to \texttt{RQST}, we have the required value of \texttt{protocol} which specifies which external call you want to make. For example, this could be defined as \texttt{graph-aave-users} which tells the DFDE to retrieve user data from AAVE using The Graph \cite{aavewebsite} \cite{thegraph2023}.
        
            Finally, we pass optional parameters which may be specific to the call. These typically fall under the values: \texttt{properties}, \texttt{headers}, \texttt{start\_date}, and \texttt{end\_date}. Both \texttt{properties} and \texttt{headers} refer to the standard URL parameters passed through a common REST API call \cite{ibm2023http}. The date parameters \texttt{start\_date} and \texttt{end\_date} are used to specify the dates of the data should it be a dated call.
        
            A sample call to the DFDE would be formatted as follows (note with no line breaks):\\
            \texttt{SRC\&\&\&RQST\&\&\&protocol\&\&\&graph-aave-users\\\&\&\&start\_date\&\&\&2022-01-01\\\&\&\&end\_date\&\&\&2023-01-01}
        
            This would send a request to \texttt{graph-aave-users} to retrieve data all user data for the year 2022.
        
        \subsubsection{Parsing the Response}
        
            Finally, the user application has to properly handle all incoming data being sent through the Socket. The format for all data passed is in JSON, with each data point being sent as an individual JSON object rather than in an array \cite{mongo2023JSON} \cite{toolsqa2023json}. This can be easily processed in most languages and then stored however the user application needs.
        
            The user application also needs to know when the data flow ends, as the connection will not automatically disconnect. Therefore, once the DFDE is done transmitting data, it will pass the character sequence \texttt{<<<end>>>} signaling the end of the data transmission.
        
            To keep the connection alive on larger data set requests, as Socket connections will automatically timeout after a short period of time, we use a "heartbeat" connection \cite{readthedocs2023timeouts} \cite{ibm2023heartbeats}. This is passed to the user at every predetermined interval of time (defaulted to 5 seconds) as the character sequence \texttt{<<<heartbeat>>>}. If the user receives this line, they can just ignore it and continue to the next line, as it is not part of the transmitted data.
    
    \subsection{Multi-tenant Architecture}
    
\begin{figure}[ht]
            \centering
            \includegraphics[scale=0.5]{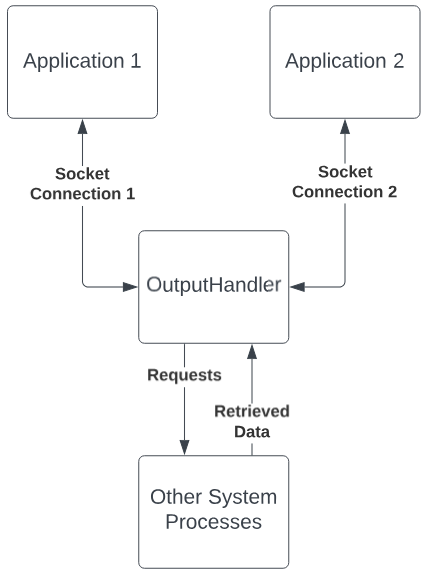}
            \caption{Socket Connection Overview}
            \label{socket-overview}
        \end{figure}

        Multi-tenant architecture refers to the usage of the same computational resources by multiple external users. In the instance of the DFDE, we utilize multi-tenant architecture by hosting a threaded ServerSocket application for users to connect to \cite{java2023serversockets} \cite{kleiman1996programming}. \cref{socket-overview} gives a visual of how the \texttt{OutputHandler} Router within the DFDE acts as a ServerSocket being able to handle multiple Socket connections at once.
    
        To keep track of these various Socket connections, we use the \texttt{destination-key} mentioned previously. The \texttt{OutputHandler} keeps a registry of each of these Socket and key pairs, routing all of the connections and requests for the DFDE.
    
        The handler also splits the storage of the consumer (incoming data stream) and producer (outgoing data stream), allowing each of them to work asynchronously. This is because only the consumer, the channel used for receiving user requests, is bound to a thread, allowing for the outward data stream to be accessed by multiple threads.

\section{Implemented Application}

    In this section, we will discuss the currently implemented application of the proposed architecture, the DFDE, and how it is working to improve the efficiency and quality of data analytics performed in a lab setting. This particular application is for the analysis of transaction data from DeFi protocols. The size and structure of the data for this application are described in \cref{sec:dataRetrieval}.

    \subsection{DeFi Data Engine}\label{sec:dfde}

        \begin{figure*}[ht]
            \centering
            \includegraphics[scale=0.6]{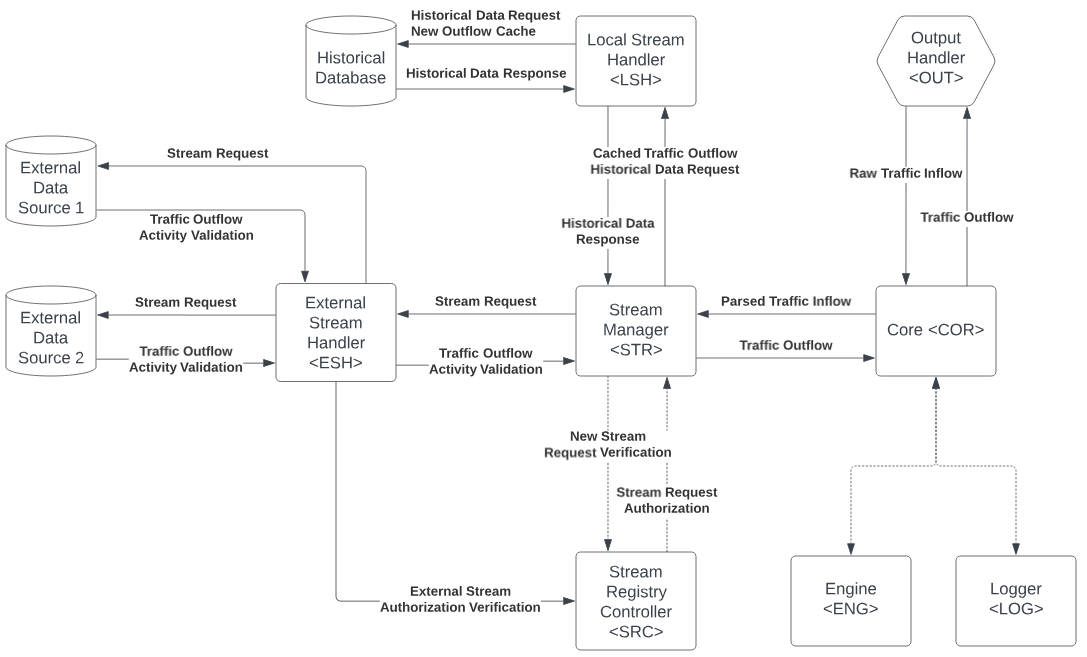}
            \caption{DeFi Data Engine Architecture}
            \label{engine-overview}
        \end{figure*}
    
        To demonstrate the efficacy of the designs discussed, we present the full architectural implementation of the DFDE. This engine utilized the Router system, internal storage mechanism, external agnostic data retrieval system, and Socket connections for communications. By using all of these systems, we can create a flow chart for the different Router connections and their interactions, as shown in \cref{engine-overview}. Each of the processes listed here is a Router, all of which contain multiple sub-processes, with the lines showing the Routers' requests between each other. 
    
        Several of these implemented Routers are imperative to the functionality of the engine. For managing data streams, the engine relies on the \texttt{LocalStreamHandler} (\texttt{LSH}), the \texttt{ExternalStreamHandler} (\texttt{ESH}), and the \texttt{StreamRegistryController} (\texttt{SRC}) to handle and process requests. When handling events such as user connections and data flows, the engine utilizes the \texttt{OutputHandler} (\texttt{OUT}). And finally, for internal communication and engine stability/information, it utilizes the \texttt{Engine} (\texttt{ENG}) and the \texttt{Logger} (\texttt{LOG}) Routers, with all other defined Routers supporting infrastructure-related processes of the system. From these brief descriptions, as well as through reference to \cref{engine-overview}, we can get a better understanding of how different data streams and requests flow throughout the system.
        
\begin{figure}
            \centering
            \includegraphics[scale=0.55]{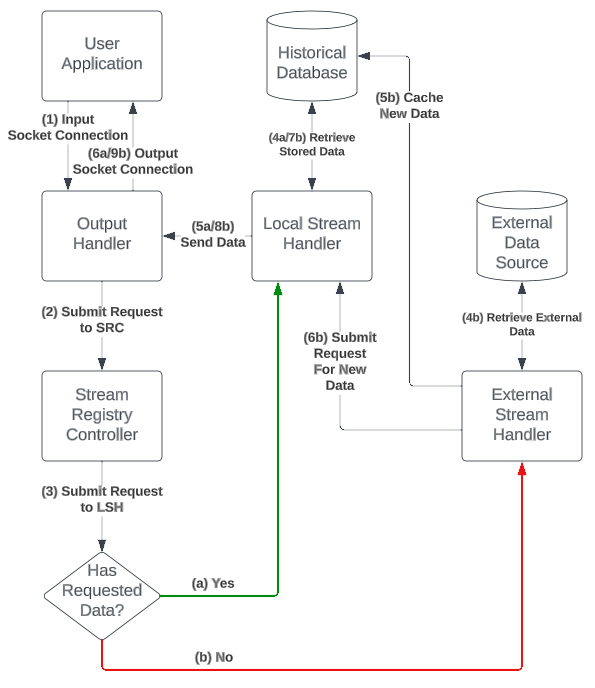}
            \caption{Sample Request Flow}
            \label{request-flow}
        \end{figure}

        The most common request is the request for stored data, where the system will either retrieve it from the historical database or from an external source. In \cref{request-flow}, we show a simplified version of the flow throughout the different Routers to execute a request.
        
    \subsection{Sample Connections}
    
        To demonstrate the language-agnostic connection capabilities of the DFDE, sample connections were developed in R-Lang \cite{rlang2016}, Python \cite{van1995python}, and Java \cite{arnold2005java}. These examples utilize the aforementioned parameters required to send a request (\texttt{protocol}, \texttt{properties}, \texttt{headers}, \texttt{startdate}, and \texttt{enddate}) and automatically instantiate the Socket connection with the server the engine is hosted on. After interpreting the request and receiving the ending statement (\texttt{<<<end>>>}), these systems will terminate the connection with the engine and output the returned data. The sample implementations for R-Lang and Python can be found on Github.
    
    \subsection{Data Retrieval}\label{sec:dataRetrieval}
    
        To further expand upon the functionality of the engine, research was conducted using it within a laboratory setting, with the primary objective of the research being the analysis of data from DeFi lending protocols as illustrated in these papers \cite{green2022defi, green2023characterizing}. 
        % AG: Below is my attempt to better explain the magnitude of data and diversity of sources used in our analysis:
        Using the DFDE for these analyses showcases the engine's ability to retrieve large datasets. The data acquired for these analyses included more than 60 million transactions from various DeFi lending protocols such as Aave~\cite{aave-whitepaperV2, aave-whitepaperV3}, Compound~\cite{compoundWhitepaper}, and MakerDAO~\cite{makerDaoWhitepaper}. Retrieving this data required the engine to connect to dozens of different tables from The Graph~\cite{thegraph2023}, API endpoints from Amberdata \cite{amberdatawebsite}, and API endpoints from DeFi Llama~\cite{defillama}. From The Graph, GraphQL~\cite{graphqlwebsite} was used to acquire raw transaction data for seven different markets of Aave. From Amberdata, a REST API connection~\cite{openapi} was used to acquire raw transaction data for Compound, MakerDAO, and one additional Aave market; blockchain address data was also acquired from Amberdata. From DeFi Llama, we pulled data for classifying cryptocurrencies as stable or non-stable coins using a REST API. Retrieving data from multiple sources aims to showcase the data-agnostic design of the engine.
    
        Aside from the retrieval of the data, preliminary processing is required to get the data in the form the researchers require. For this, the sample file of \texttt{GetTransactions.Rmd} is created, which handles all internal processing of the data so that it can be loaded easily with a simple function call \cite{flynn2023rtransactions}. It makes use of 13 sample requests, showcasing the data-agnostic designs of the engine \cite{flynn2023requests}.
    
        Once the data is all retrieved and processed, it will be presented to the user in a list of two objects. The first object within the list is \texttt{response} which details the outcome of the call to the engine. This contains the \texttt{code} (code corresponding to the outcome of the call), \texttt{message} (accompanying message to the response code), and \texttt{data} (any data that may be returned from the call). The second object is the data frame previously mentioned under the key \texttt{df}, which contains all returned and processed data.
    
    \subsection{Data Analytics}
    
        Two sample plots are used to show how the retrieved processed data can be used in research. Each only uses data retrieved from the engine with no external supporting data. 
    
\begin{figure}
            \centering
            \includegraphics[scale=0.35]{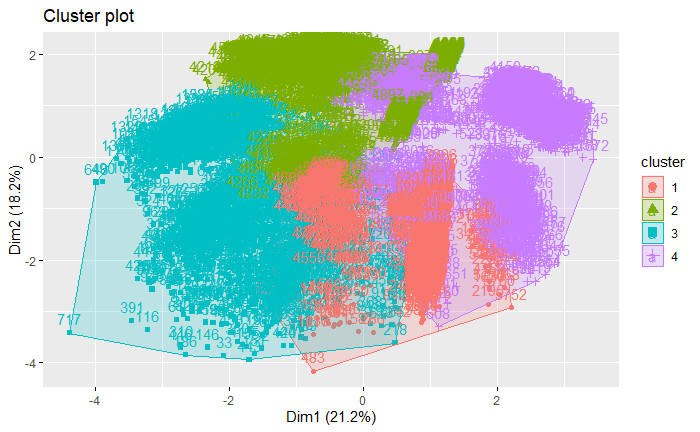}
            \caption{User Clustering Plot}
            \label{cluster-plot}
        \end{figure}

        The first plot, shown in \cref{cluster-plot}, is a k-means clustering plot assigning different clusters to users of the AAVE lending protocol based on their skill level and risk profile derived from their transaction history \cite{likas2003global, green2022defi}. From this chart, we can see this clustering and each of the different users found within the lending transaction data being placed into one of four categories. 
    
\begin{figure}
            \centering
            \includegraphics[scale=0.35]{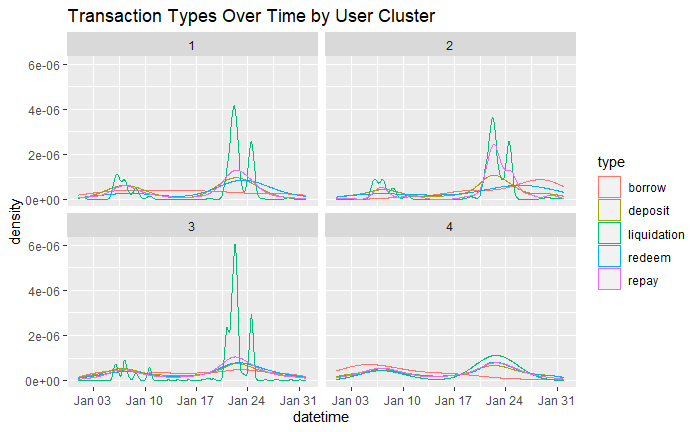}
            \caption{Transaction Type Density Plot}
            \label{density-plot}
        \end{figure}

        The second plot, shown in \cref{density-plot}, is a density plot, which informs the researcher how many of each transaction type occurred during a certain time period based on the clusters found in \cref{cluster-plot}.
        
        Although these plots are demonstrative visual overviews of the data retrieved, more meaningful points can certainly be extrapolated with more time based on the given researchers' needs.

\section{Conclusion}

    To further improve this application of the proposed systems, developers and researchers can make use of further external communication generalizations. Outside of the previously discussed REST API generalized connections already implemented, developers can potentially integrate connections such as live connections to external Websocket applications, generalized connections to other query languages such as SQL and NoSQL databases, and further improve the efficiency of the internal designs \cite{stonebraker2010sql}.

    Consequently, we have proposed a language and data agnostic system called the "DeFi Data Engine" which aims to generalize data retrieval and storage from multiple external sources for easy usage. First, we explored the internal architecture of the system, analyzing its performance capabilities, how it can be used to scale systems, and the embedded data agnostic communication architecture. Then, we discussed the internal and external communication processes developed to support the DFDE and how the implementation can handle the different formatting of data from these sources. Finally we discuss the DFDE being used as an application in classroom and lab settings, as well as the results and papers that were derived from its usage. 

\section{Acknowledgements}

    The authors acknowledge the support from NSF IUCRC CRAFT center research grants (CRAFT Grants \#22003, \#22006) for this research. The opinions expressed in this publication and its accompanying code base do not necessarily represent the views of NSF IUCRC CRAFT.
    This work was supported by the Rensselaer Institute for Data Exploration and Applications (IDEA).\footnote{https://idea.rpi.edu/} We also thank Amberdata for supporting us with access to their data.

\bibliographystyle{IEEEtran}
\bibliography{bibi}
\vspace{12pt}
\end{document}